\documentclass{PoS}
\usepackage{amsmath,amssymb,epsfig,fleqn,url,color,here,graphics,graphicx,rotating,multirow,wrapfig,enumerate}
\usepackage{etex}
\usepackage{caption}
\usepackage{subcaption}
\usepackage{pbox}
\usepackage{dcolumn}
\usepackage{bigstrut}
\usepackage{soul,xcolor}
\usepackage{lineno, xcolor}

\usepackage[section]{placeins}

\newcommand{\lf}{\left}
\newcommand{\rg}{\right}

\newcommand{\phiCS}{\varphi _{CS}}
\newcommand{\phiS}{\varphi _S}
\newcommand{\thCS}{\theta _{CS}}
\newcommand{\gvc}{GeV/$c$}

\newcommand{\gvcw}{GeV/$c^2$}

\newcommand{\gvct}{GeV/$c^{ 2}$ }
\newcommand{\Minv}{$M_{\mu\mu}$}
\newcommand{\Jpsi}{J$/\psi\,$}

\setstcolor{red}
\title{First measurement of transverse-spin-dependent azimuthal asymmetries in the Drell-Yan process}

\ShortTitle{Measurement of LSAs in SIDIS at COMPASS experiment}

\author{\speaker{Bakur Parsamyan}\\
        CERN, University of Turin and INFN section of Turin\\
        E-mail: \email{bakur.parsamyan@cern.ch}}

\abstract{The COMPASS experiment at CERN, as part of  its programme addresses the exploration of the transverse spin structure of the nucleon by measuring spin (in)dependent azimuthal asymmetries in semi-inclusive DIS and, recently, also in Drell-Yan processes. Between 2002 and 2010 COMPASS performed a series of SIDIS measurements, using a longitudinally polarized muon beam impinging on transversely polarized $^6LiD$ or $NH_3$ targets. Drell-Yan measurements with a $\pi^-$ beam interacting with a transversely polarized $NH_3$ target started with the 2015 run and will be continued in 2018.
In this Letter the first measurement of transverse-spin-dependent azimuthal asymmetries in the pion-induced Drell-Yan process is reported. Measured asymmetries giving access to different transverse-momentum-dependent (TMD) parton distribution functions (PDFs) are extracted using dimuon events with invariant mass between 4.3~\gvcw\, and 8.5~\gvcw.
A recent COMPASS SIDIS measurement was obtained at a hard scale comparable to that of these DY results.
This provides a unique possibility to test predicted in QCD sign change of the Sivers TMD PDF and other (pseudo-)universal features of transvers momentum dependent parton distribution functions.}

\FullConference{XXV International Workshop on Deep-Inelastic Scattering and Related Subjects\\
		3-7 April 2017\\
		University of Birmingham, UK}
\begin{document}
\section{Introduction}
\label{sec:intro}
Within Leading Order (LO) QCD parton model approach the polarized nucleon is described by six time reversal even and two time reversal odd \textit{twist-2} quark transverse momentum dependent (TMD) parton distribution functions (PDFs). The TMD PDFs are universal, process-independent functions\footnote{QCD \textit{generalized universality}: time-reversal modified process-independence of TMD PDFs}~\cite{Collins:2011zzd} describing longitudinal and transverse momenta distributions of partons and their correlations with nucleon and quark spins. Such correlations induce azimuthal modulations (asymmetries) in the cross sections of SIDIS ($\ell \,N \rightarrow \ell^\prime \,h \, X$, semi-inclusive hadron production in deep-inelastic lepton-nucleon scattering) and of Drell-Yan process ($h \, N \rightarrow \ell\,\bar{\ell}\, X$, massive lepton-pair production in hadron-nucleon collisions). Applying the TMD factorization theorems~\cite{Collins:2011zzd} allows one to express the asymmetries raising in DY and SIDIS cross sections in terms of convolutions of perturbatively calculable hard-scattering parton cross sections, hard-scale dependent TMD PDFs and (for SIDIS) parton fragmentation functions (FFs). The hard scale $Q$ in SIDIS is given by the square root of the virtuality of the photon exchanged in the DIS process and in DY by the invariant mass of the produced lepton pair.

Measurements and following study of the spin (in)dependent azimuthal effects in SIDIS and Drell-Yan is a powerful method to access TMD distribution functions of the nucleon, which in past decades became a priority direction in experimental and theoretical high-energy physics, for recent reviews see \textit{e.g.} Refs.~\cite{Aidala:2012mv,Boglione:2015zyc}.
The ultimate goal is to measure experimentally with high precision all possible spin-effects with both SIDIS and Drell-Yan reactions at different energies and perform global multi-differential analysis of obtained results to extract all spin-dependent distribution functions.

\begin{wrapfigure}{r}{6cm}
\centering
\includegraphics[width=0.39\textwidth]{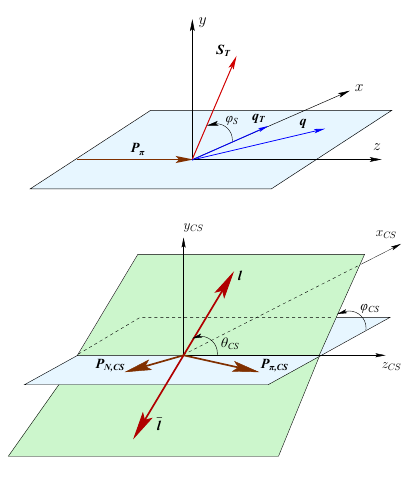}
\caption{The target rest (top) and the Collins-Soper frames (bottom).
}
\label{fig:DYframe}
\end{wrapfigure}

When the polarizations of the produced leptons are summed over, the general expression for the cross section of pion induced DY lepton-pair production off a transversely polarized nucleon comprises one polar asymmetry, two \textit{unpolarized} and five target transverse-spin-dependent azimuthal asymmetries (TSAs). Adopting general notations and conventions of Refs.~\cite{Arnold:2008kf,Gautheron:2010wva}, the model-independent differential cross section can be written as follows:
%
%
\begin{eqnarray}\label{eq:DY_xsecLO}
  \hspace*{0cm}\frac{d\sigma}{dq^{4}d\Omega} &\propto& \hat{\sigma}_{U}\bigg\{ 1 + D_{[\sin{2\thCS}]} A_U^{\cos {\phiCS}}\cos {\phiCS}+
  D_{[\sin^{2}\thCS]} A_U^{\cos 2{\phiCS}}\cos 2{\phiCS}\\ \nonumber
   &&\hspace*{0.9cm} +  {S_T}\Big[ D_{[1+\cos^{2}\thCS]}A_T^{\sin {\phiS}}\sin {\phiS} \\ \nonumber
   &&\hspace*{1.4cm} + D_{[\sin^{2}\thCS]}
       \lf( A_T^{\sin( {2{\phiCS} - {\phiS}}    )}\sin ( {2{\phiCS} - {\phiS}} )
   { + A_T^{\sin \lf( {2{\phiCS} + {\phiS}} \rg)}}\sin \lf( {2{\phiCS} + {\phiS}} \rg) \rg)\\ \nonumber
   &&\hspace*{1.4cm} + D_{[\sin{2\thCS}]}
       \lf( A_T^{\sin( {{\phiCS} - {\phiS}}    )}\sin ( {{\phiCS} - {\phiS}} )
   { + A_T^{\sin \lf( {{\phiCS} + {\phiS}} \rg)}}\sin \lf( {{\phiCS} + {\phiS}} \rg) \rg)
   \Big] \bigg\},
\end{eqnarray}
%
%
Here, $q$ is the four-momentum of the exchanged virtual photon, $F^{1}_{U}$, $F^{2}_{U}$ are the polarization and azimuth-independent structure functions and $\hat{\sigma}_{U} =  \lf({F^{1}_{U}}+{F^{2}_{U}}\rg)\lf(1 + \lambda{{\cos}^2}\thCS \rg)$, with $\lambda$ being the polar angle asymmetry, given as $\lambda=\lf({F^{1}_{U}}-{F^{2}_{U}}\rg)/\lf({F^{1}_{U}}+{F^{2}_{U}}\rg)$. The subscript ($U$)$T$ denotes target transverse polarization (in)dependence. In analogy to SIDIS, the virtual-photon depolarization factors are defined as $D_{[f(\thCS)]}=f(\thCS)/\lf(1 + \lambda{{\cos }^2}\thCS \rg)$ with $f(\thCS)$ being equal either to $\sin2\thCS$, or $\sin^2\thCS$, or to $1+\cos^2\thCS$. The angles $\phiCS$, $\thCS$ and $\Omega$, the solid angle of the lepton, are defined in the Collins-Soper frame following the considerations of Refs.~\cite{Arnold:2008kf,Gautheron:2010wva}, and $\phiS$ is the azimuthal angle of the direction of the nucleon polarization in the target rest frame, see Fig.~\ref{fig:DYframe}.

The asymmetries $A_{(U)T}^{w}$ in Eq.~\ref{eq:DY_xsecLO} are defined as amplitudes of a given azimuthal modulation $w=w(\phiS,\phiCS)$, divided by the spin and azimuth-independent part of the DY cross section and the corresponding depolarization factor.
One spin-independent asymmetry and three out of five TSAs enter at leading order of  perturbative QCD and can be described by contributions from only \textit{twist-2} TMD PDFs. In DY lepton-pair production with a transversely polarized nucleon in the initial state, the $A_U^{\cos2\phiCS}$ asymmetry is related to the convolution of nucleon and pion Boer-Mulders TMD PDFs, ($h_{1}^{\perp}$ and $h_{1,\pi}^{\perp}$, correspondingly).
The TSA $A_T^{\sin\phiS}$ is related to the nucleon Sivers TMD PDFs ($f_{1T}^\perp$) convoluted with the unpolarized pion TMD PDFs ($f_{1,\pi}$). Here, following similar SIDIS conventions, \textit{twist-3} contribution to the Sivers TSA is neglected\footnote{the $F^{2}_{UT}$ structure function is assumed to be zero~\cite{Gautheron:2010wva}}. Analogously, within the \textit{twist-2} approximation of LO pQCD, $F^{2}_{U}=0$ and therefore $\lambda=1$.
The other two \textit{twist-2} TSAs, $A_T^{\sin(2\phiCS-\phiS)}$ and $A_T^{\sin(2\phiCS+\phiS)}$, are related to convolutions of the Boer-Mulders TMD PDFs ($h_{1,\pi}^{\perp}$) of the pion with the nucleon TMD PDFs transversity ($h_1$) and pretzelosity ($h_{1T}^\perp$), respectively~\cite{Arnold:2008kf,Bacchetta:2006tn}. Remaining three azimuthal asymmetries, namely $A_U^{\cos\phiCS}$, $A_T^{\sin(\phiCS-\phiS)}$ and $A_T^{\sin(\phiCS+\phiS)}$ are \textit{subleading-twist} structures and are expected to vanish at LO.

All three aforementioned \textit{twist-2} nucleon TMD PDFs related to LO DY TSAs induce analogous \textit{twist-2} TSAs in the general expression for the cross section of unpolarized-hadron production in SIDIS of leptons off transversely polarized nucleons~\cite{Arnold:2008kf,Bacchetta:2006tn,Kotzinian:1994dv}. The $A_{UT}^{sin(\phi_h-\phi_S)}$ (Sivers) and $A_{UT}^{sin(\phi_h+\phi_S)}$ (Collins) SIDIS TSAs are the most studied ones. Corresponding structure functions are given as convolutions of Sivers PDF with ordinary fragmentation function ($D_{1}$) and transversity PDF with Collins FF ($H_{1}^{\perp}$), respectively. The $A_{UT}^{\sin(3\phi_h -\phi_S )}$ SIDIS asymmetry is related to pretzelosity PDF convoluted with Collins FF.
These TSAs were studied by the HERMES and COMPASS experiments, see Refs.~\cite{Adolph:2012sp,Adolph:2012sn,Airapetian:2009ae,Adolph:2016dvl,Parsamyan:2013ug}.

The Sivers function~\cite{Sivers:1989cc} plays an important role among the TMD PDFs. It describes the left-right asymmetry in the distribution of unpolarized partons in the nucleon with respect to the plane spanned by the momentum and spin vectors of the nucleon.
One of the recent significant theoretical advances in the TMD framework of QCD is the prediction that the two naively time-reversal odd TMD PDFs, \textit{i.e.} the quark Sivers functions $f_{1T}^\perp$ and Boer-Mulders functions $h_{1}^\perp$, have opposite sign when measured in SIDIS on the one hand, and in DY or $W$/$Z$-boson production on the other.
~\cite{Collins:2002kn}.
The experimental test of this fundamental prediction, which is a direct consequence of QCD gauge invariance, is a major challenge in hadron physics. In contrast to the  Sivers and Boer-Mulders functions, transversity and pretzelosity TMD PDFs are predicted to be genuinely universal, \textit{i.e.} they do not change sign between SIDIS and DY~\cite{Collins:2011zzd}, which is yet another fundamental QCD prediction to be explored.

Non-zero quark Sivers TMD PDFs have been extracted from SIDIS single-differential results of HERMES~\cite{Airapetian:2009ae}, COMPASS~\cite{Adolph:2012sp} and JLab~\cite{Qian:2011py} using both collinear and TMD evolution approaches~\cite{Anselmino:2016uie,Echevarria:2014xaa,Sun:2013hua}.
The first measurement of the Sivers effect in $W$ and $Z$-boson production in collisions of transversely polarized protons at RHIC was reported by the STAR collaboration~\cite{Adamczyk:2015gyk};
the hard scales of these measurements is $Q\approx 80$\;\gvc\; and $90$\;\gvc. It is quite different from the one explored in fixed-target experiments where $Q$ ranges approximately between $1$\;\gvc\; and $9$\;\gvc. Hence it is not excluded that TMD evolution effects may be sizable when describing the STAR results using Sivers TMD PDFs extracted from fixed-target SIDIS results.

\begin{wrapfigure}{r}{5cm}
\centering
\includegraphics[width=0.33\textwidth]{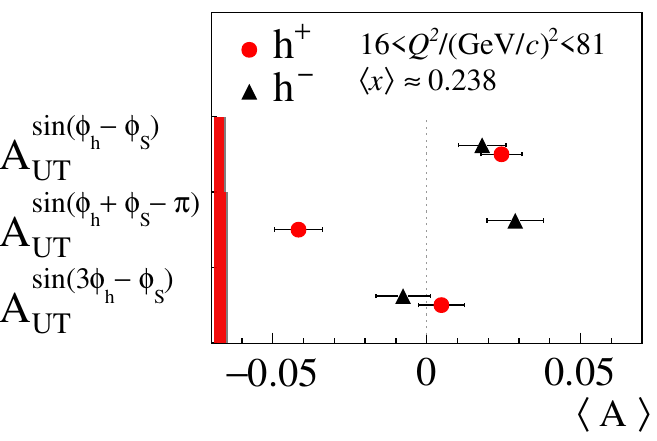}
\caption{COMPASS results for proton SIDIS TSAs in the range 4 \gvc$\;<Q<$ 9 \gvc.}
\label{fig:A8in1}
\end{wrapfigure}
The COMPASS experiment at CERN~\cite{Gautheron:2010wva} has the unique capability to explore the transverse-spin structure of the nucleon in a similar kinematic region by two alternative experimental approaches, \textit{i.e.} SIDIS and DY, using mostly the same setup.
This offers the opportunity of minimizing uncertainties of TMD evolution in the comparison of the Sivers TMD PDFs when extracted from these two measurements to test the opposite-sign prediction by QCD.

Recently, COMPASS published the first multi-differential results of the TSAs, which were extracted from SIDIS data at four different hard scales~\cite{Adolph:2016dvl}.
The three aforementioned TSAs that are extracted from COMPASS SIDIS data for the range 4 \gvc$\;<Q<$ 9 \gvc, are shown in Fig.~\ref{fig:A8in1} after averaging over all other kinematic dependences. This hard scale range is very similar to the one used in this Letter to analyze the DY process.
The Sivers asymmetry for positive hadrons was found to be above zero by 3.2 standard deviations of the total experimental accuracy. The Collins asymmetry is determined with even better statistical precision. The amplitude has opposite sign for positive and negative hadrons which is attributed to the peculiarities of Collins FF~\cite{Adolph:2012sn}. The pretzelosity asymmetry in SIDIS is found to be compatible with zero, which can be related to kinematic suppression-factors entering in the corresponding structure function~\cite{Bacchetta:2006tn,Parsamyan:2013ug}.

COMPASS results for three \textit{twist-2} Drell-Yan TSAs were published in Ref.~\cite{Aghasyan:2017jop}. In this Letter COMPASS results for all five proton transverse-spin-dependent asymmetries in the pion-induced Drell-Yan process are presented for the first time. Obtained results are given in various kinematic representations in order to provide more detailed input for relevant studies of involved TMD PDFs.

\section{Data analysis}
\label{sec:data_analysis}
\begin{wrapfigure}{r}{7cm}
\centering
\includegraphics[width=0.39\textwidth]{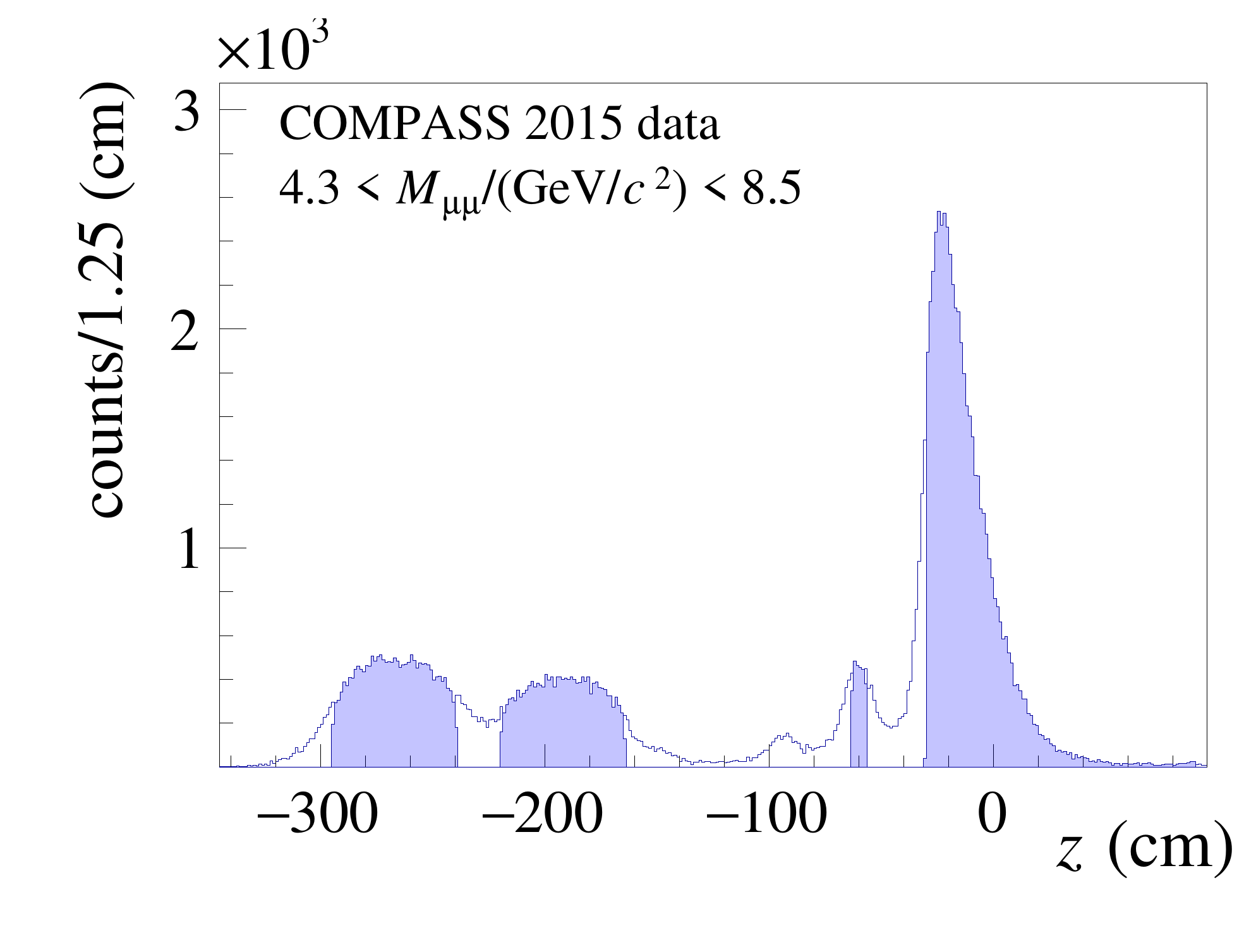}
\caption{Distribution of dimuon production vertices in the target region.}
\label{fig:PVz}
\end{wrapfigure}

The analysis presented in this Letter is based on Drell-Yan data collected by COMPASS in the year 2015 using essentially the same spectrometer as it was used during SIDIS data taking in previous years~\cite{Gautheron:2010wva}.
For this measurement, the 190 \gvc\; $\pi^{-}$ beam with an average intensity of $0.6\times10^{8}$ s$^{-1}$ from the CERN SPS - M2 beamline was scattered off the COMPASS transversely polarized NH$_3$ target.
The polarized target, placed in a 0.6~T dipole magnet, consisted of two longitudinally aligned cylindrical cells of 55 cm length and 4~cm in diameter, separated
by a 20~cm gap.
A 240 cm long structure made mostly of alumina with a tungsten core, placed downstream of the target, acted as hadron absorber and beam dump. In addition, a 7cm long aluminum plug dedicated to the unpolarized DY-measurements was installed inside the absorber upstream of the tungsten beam plug. Approximately $60\%$ of the beam did not interact in the polarized target and was dumped into the tungsten plug. The distribution of dimuon production vertices in the target region is shown in Fig.~\ref{fig:PVz} with indicated target cell, aluminum plug and tungsten core positions.

\begin{figure}[h!]
\centering
\includegraphics[width=1.0\textwidth]{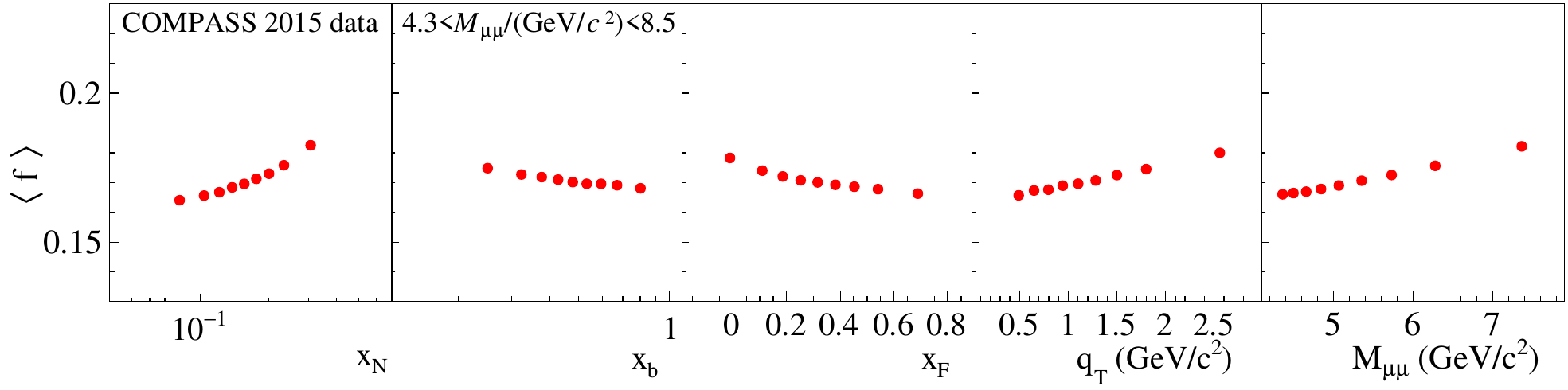}
\caption{Kinematic dependence of the dilution factor.}
\label{fig:f_x}
\end{figure}

The average uncertainty in the reconstructed vertex position along the beam line is about $10$~cm, which leads to a marginal (below 1\%)  migration of reconstructed events from one target cell to the other.
The two cells were polarized vertically in opposite directions, so that data with both spin orientations were recorded simultaneously. In order to compensate for acceptance effects, the polarization was reversed every two weeks.
The entire data-taking time of 18 weeks was divided into nine periods, each consisting of two consecutive weeks with opposite target polarizations.
The proton polarization had a relaxation time of about 1000 hours, which was measured for each target cell in each data taking period. Average proton polarization was measured to be around $\langle P_T\rangle\approx0.73$. The dilution factor, $f$, accounting for the fraction of polarizable nucleons in the target and the migration of reconstructed events from one target cell to the other or from unpolarized medium into the cells, is estimated to be $\langle f\rangle\approx 0.18$.
Kinematic dependence of the dilution factor is shown in Fig.~\ref{fig:f_x}.

Outgoing charged particles were detected by a system of tracking detectors in the two-stage spectrometer and their momenta are determined by means of two large-aperture dipole magnets. In each stage, muon identification was accomplished by a system of muon filters.
The trigger required the hit pattern of
several hodoscope planes to be consistent with at least two muon candidates originating from the target region. For any pair of candidates either both have to be detected in the first stage of the spectrometer ($25 < \theta_{\mu} < 160$~mrad), or one in the first and the other in the second stage ($8 < \theta_{\mu} < 45$~mrad).

In the data analysis, the selection of events requires a production vertex located within the polarized-target volume, with one incoming pion beam track and at least two oppositely charged outgoing particles that are consistent with the muon hypothesis, ensured by the requirement of 30 radiation lengths to be crossed along the spectrometer. The quality and timing information for the tracks is also verified.
The dimuon transverse momentum $q_T$ is required to be above $0.4$ \gvc\; in order to obtain sufficient resolution in angular variables.

For various Drell-Yan studies carried out at COMPASS, it is convenient to disentangle four \Minv dimuon invariant mass-ranges ~\cite{Gautheron:2010wva}:
\begin{enumerate}[i)]
  \item 1 \gvct$<M_{\mu\mu}<$ 2 \gvct: "low mass" range, many background processes contribute;
  \vspace{-0.25cm}
  \item 2 \gvct$<M_{\mu\mu}<$ 2.5 \gvct: "intermediate mass" range;
  \vspace{-0.25cm}
  \item 2.5 \gvct$<M_{\mu\mu}<$ 4.3 \gvct: "charmonium (\Jpsi and $\psi'$) mass range";
  \vspace{-0.25cm}
  \item 4.3 \gvct$<M_{\mu\mu}<$ 8.5 \gvct: "high mass" range, background processes are suppressed.
\end{enumerate}
%

\begin{wrapfigure}{r}{7cm}
\centering
\includegraphics[width=0.39\textwidth]{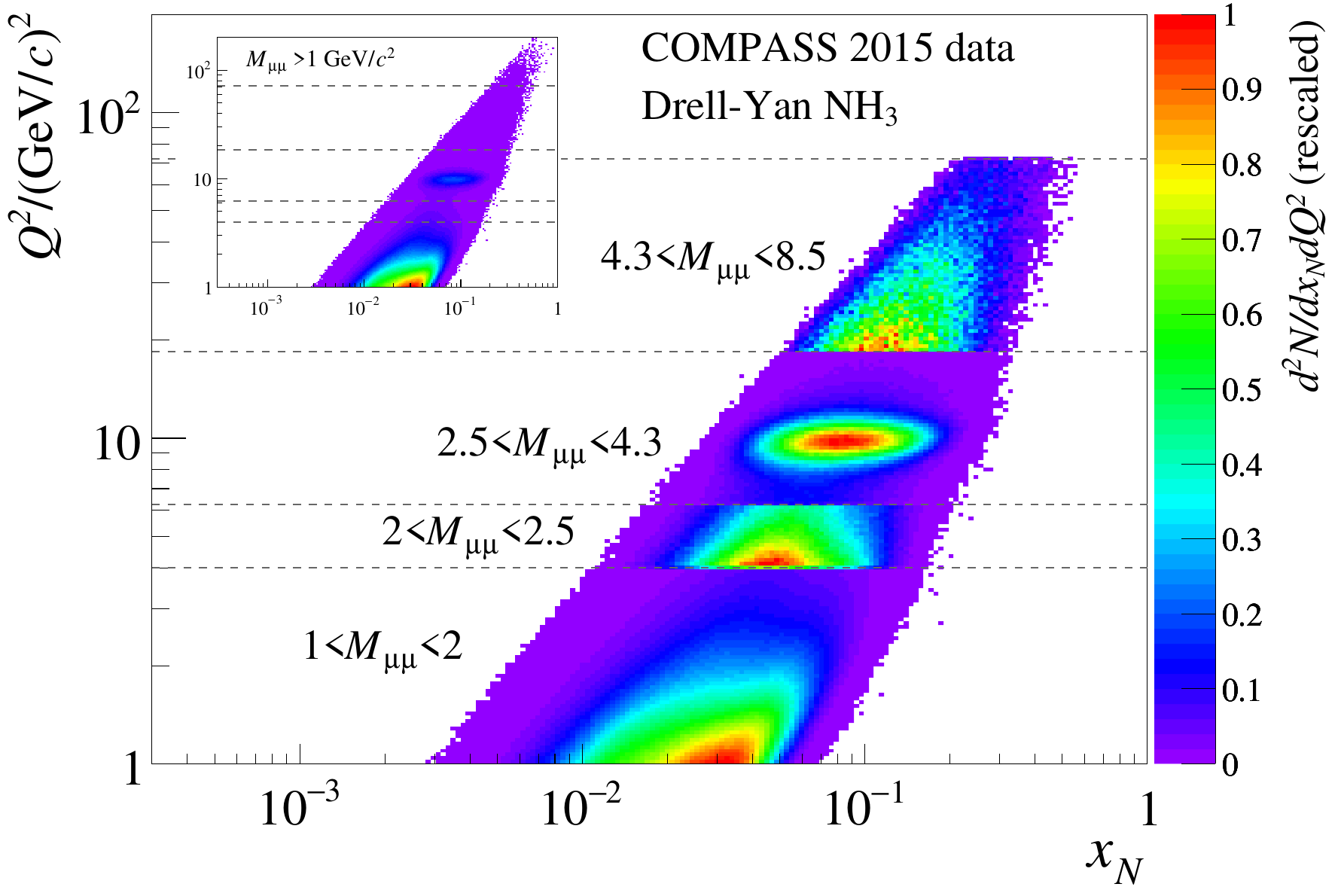}
\caption{the two-dimensional $(x_{N},Q^2)$ distribution.}
\label{fig:xQ2}
\end{wrapfigure}

In Fig.~\ref{fig:xQ2} the two-dimensional $(x_{N},Q^2)$ distributions for dimuons produced in polarized target are shown for the entire $(x_{N},Q^2)$ range (upper left insert) and separately for each $Q^2$ range. In each cell, the distribution is normalized to have a maximum value equal to one.
Range iv) is particularly suited to study the predicted sign change of the Sivers TMD PDF when comparing SIDIS and DY results. First, this range best fulfils the requirement of TMD factorisation that the transverse momentum of the hadron in SIDIS or of the muon pair in DY has to be much smaller than $Q$.
Secondly, both SIDIS and DY cross sections for a proton target are dominated by the contribution of $u$-quark nucleon TMD PDFs in the valence region, where the extracted Sivers TMD PDF reaches its maximum~\cite{Anselmino:2016uie}.
In this Letter only results obtained for range iv) are presented.

\begin{wrapfigure}{r}{7cm}
\centering
\includegraphics[width=0.39\textwidth]{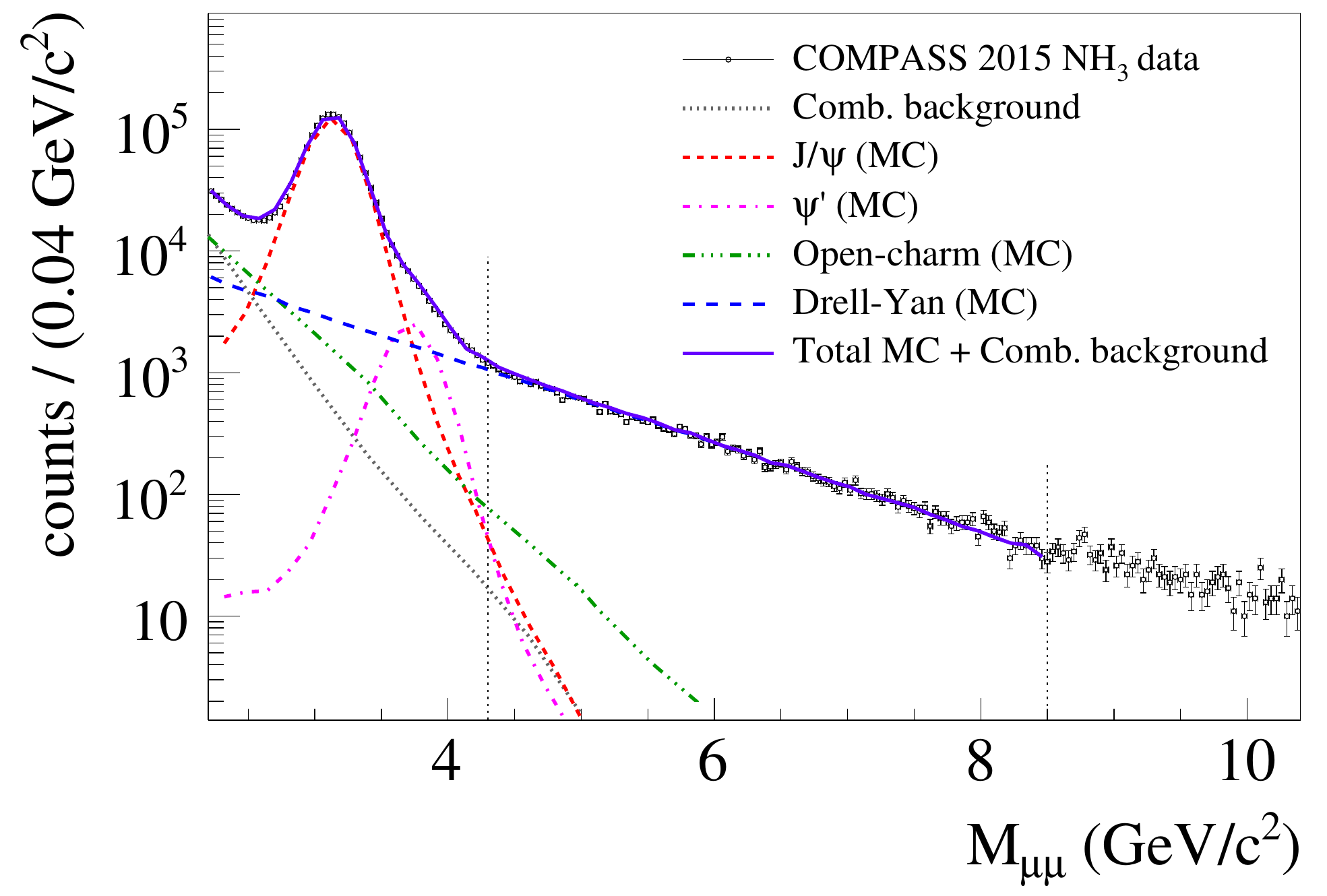}
\caption{The \Minv distribution.}
\label{fig:M}
\end{wrapfigure}

The reconstructed mass spectrum of events passing all analysis requirements is shown in black in Fig.~\ref{fig:M}.
The combinatorial background originating from the decays of pions and kaons produced in the target is evaluated using like-sign dimuon events from real data and shown in grey (dotted).
Further contributions to the dimuon spectrum are evaluated with a Monte-Carlo (MC) simulation, and their relative weights are obtained by a fit to the data.
The Drell-Yan contribution is shown in blue (long dashed). The background contributions originate from charmonia, shown in red (dashed) and magenta (dot-dashed), and semi-muonic open-charm decays shown in green (double dot dashed).
The sum of all contributions, shown in violet, describes the experimental data well. The J/$\psi$ peak is clearly visible with a shoulder from the $\psi$(2S) resonance.  The mass range iv) used in this analysis is defined by the requirement 4.3~\gvcw\,$ < M_{\mu\mu} < $\,8.5~\gvcw\,, where the upper limit avoids the contribution of $\Upsilon$-resonances. In this range, the overall background contribution is estimated to be below 4\%.

\begin{figure}[h!]
\centering
\includegraphics[width=0.37\textwidth]{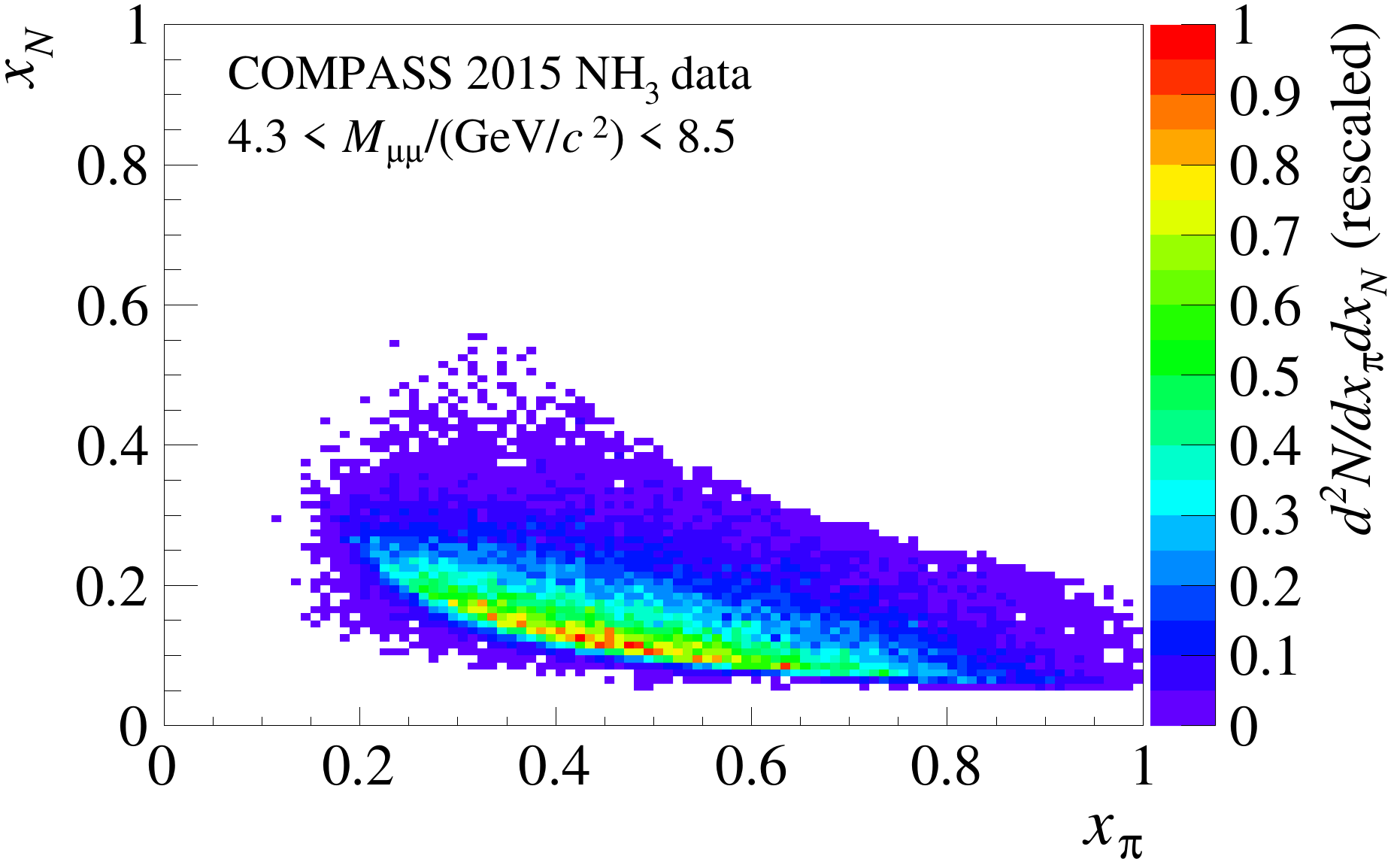}
\includegraphics[width=0.30\textwidth]{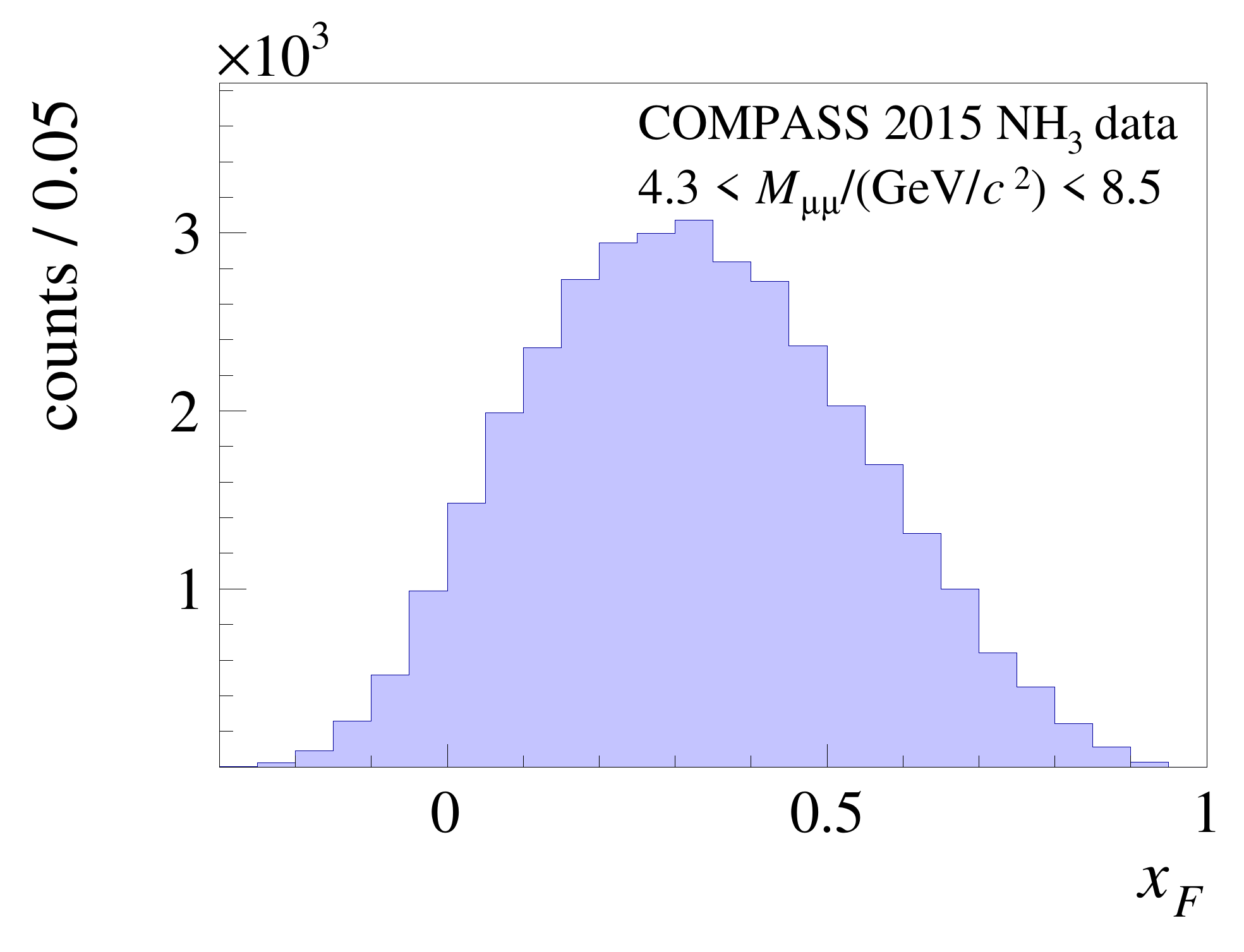}
\includegraphics[width=0.30\textwidth]{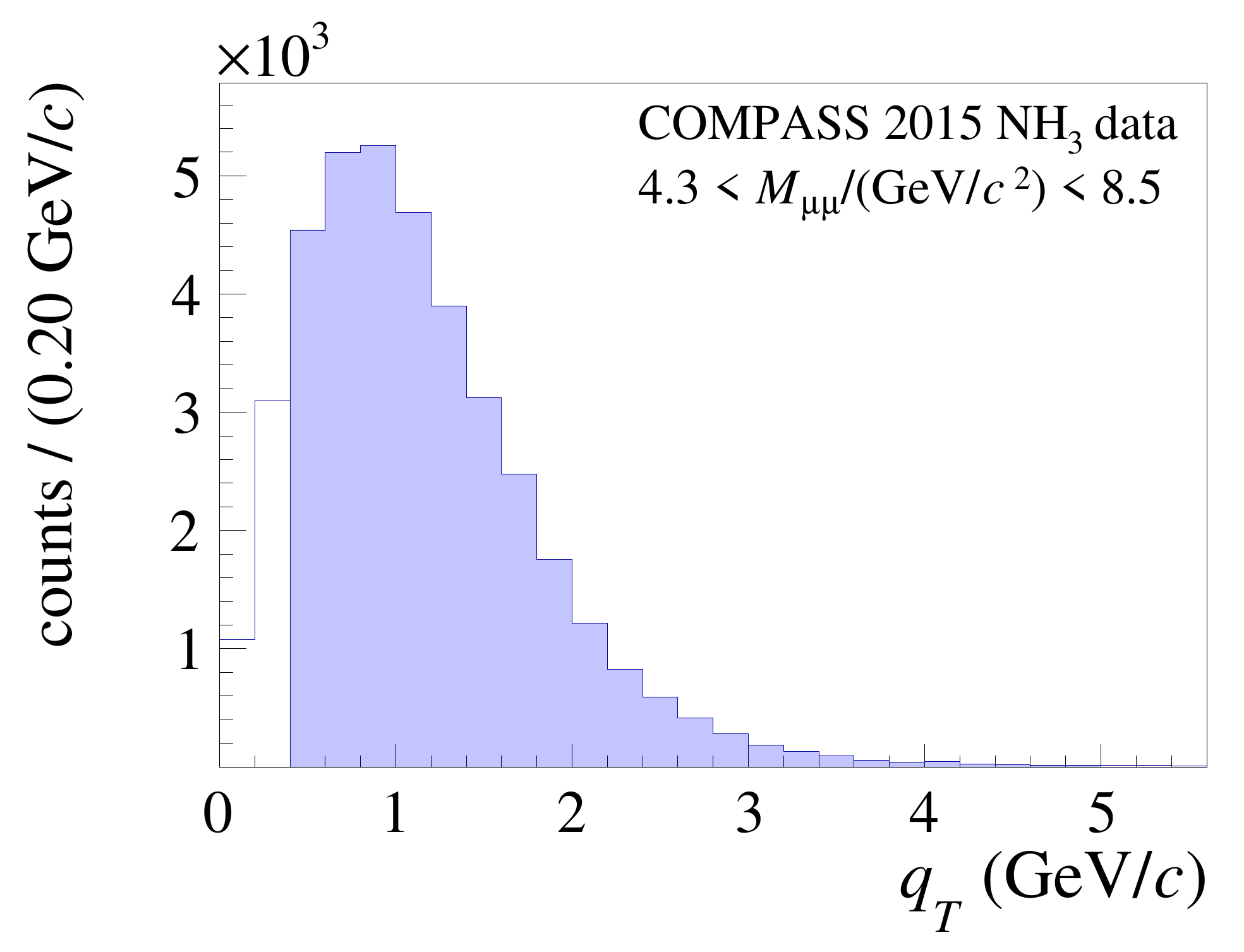}
\caption{The two-dimensional $(x_{\pi},x_{N})$ distribution of the selected high mass dimuons (left). The $x_{F}$ distribution (center) and $q_T$ distribution (right)
of the selected high mass dimuons.}
\label{fig:xqT}
\end{figure}

The two-dimensional distribution of the Bjorken scaling variables of pion and nucleon, $x_{\pi}$ and $x_{N}$, for this mass range is presented in Fig.~\ref{fig:xqT} (left panel). The figure shows that the kinematic phase space explored by the COMPASS spectrometer matches the valence region in $x_{\pi}$ and $x_{N}$. In this region, the DY cross section for a proton target is dominated by the contribution of nucleon $u$-quark and pion $\bar{u}$-quark TMD PDFs.

The distributions of the dimuon Feynman variable $x_{F}$ and the dimuon transverse momentum $q_{T}$ are presented in Fig.~\ref{fig:xqT} (central and right panels). The corresponding mean values of the kinematic variables are: $\langle x_{N} \rangle=0.17$,
$\langle x_{\pi} \rangle=0.50$,
$\langle x_{F} \rangle=0.33$,
$\langle q_{T} \rangle=1.2$~\gvc\, and
$\langle M_{\mu\mu} \rangle=5.3$~\gvcw.

After all selection criteria about $35\times10^3$ dimuons remain for the analysis.
The three TSAs presented in this Letter are extracted period by period from the number of dimuons produced in each cell for the two directions of the target polarization.
The double-cell target configuration in conjunction with the periodic polarization reversal allows for the simultaneous measurement of azimuthal asymmetries for both target spin orientations.
Using an extended Unbinned Maximum Likelihood estimator, all five TSAs are fitted simultaneously together with their correlation matrices. In this approach, flux and acceptance-dependent systematic uncertainties are minimized~\cite{Adolph:2016dvl}.
The final asymmetries are obtained by averaging the results of the nine periods.
The asymmetries are evaluated in kinematic bins of $x_{N}$, $x_{\pi}$, $x_{F}$ or $q_{T}$, while always integrating over all the other variables.
%
\begin{figure}[h!]
\centering
\includegraphics[width=1.0\textwidth]{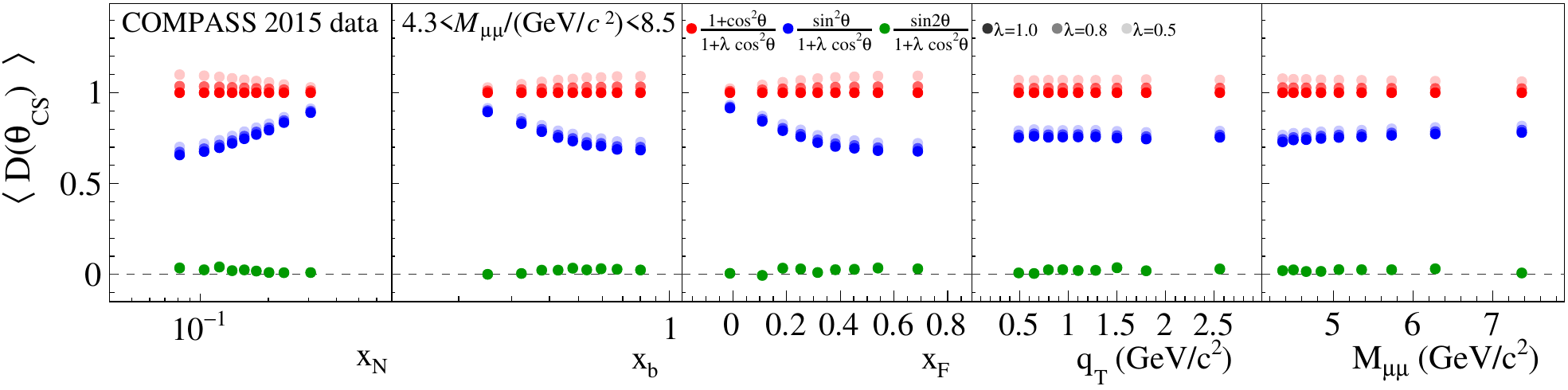}
\caption{Kinematic dependence of the depolarization factors and impact of different values of $\lambda$.}
\label{fig:Dth}
\end{figure}
The dilution factor $f$ and the depolarization factors entering the definition of TSAs are calculated on an event-by-event basis and used to weight the asymmetries. For the magnitude of the target polarization $P_T$, an average value is used for each data taking period in order to avoid possible systematic bias. In the evaluation of the depolarization factors, the approximation $\lambda=1$ is used. Known deviations from this assumption with $\lambda$ ranging between 0.5 and 1~\cite{Falciano:1986wk} decrease the normalization factor by at most $5\%$, see Fig.~\ref{fig:Dth}.

The TSAs resulting from different periods are checked for possible systematic effects. The largest systematic uncertainty is due to possible residual variations of experimental conditions within a given period. They are quantified by evaluating various types of false asymmetries in a similar way as described in Refs.~\cite{Adolph:2012sp,Adolph:2012sn}. The systematic point-to-point uncertainties are found to be about 0.7 times the statistical uncertainties and are indicated in the plots by color bands.
The normalization uncertainties originating from the uncertainties on target polarization ($5\%$) and dilution factor ($8\%$) are not included in the quoted systematic uncertainties.

\section{Results and Discussion}
\label{sec:results}
\begin{figure}[h!]
\centering
\begin{subfigure}{0.7\textwidth}
\includegraphics[width=0.95\textwidth]{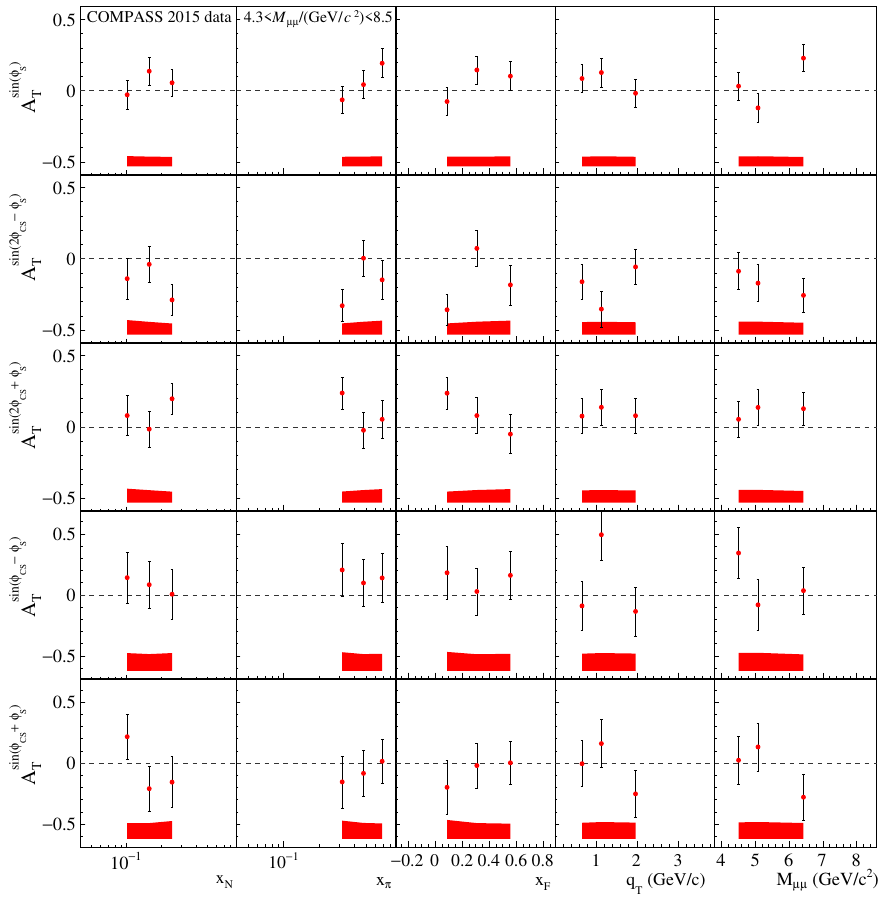}
\end{subfigure}%
\begin{subfigure}{0.3\textwidth}
\includegraphics[width=0.95\textwidth]{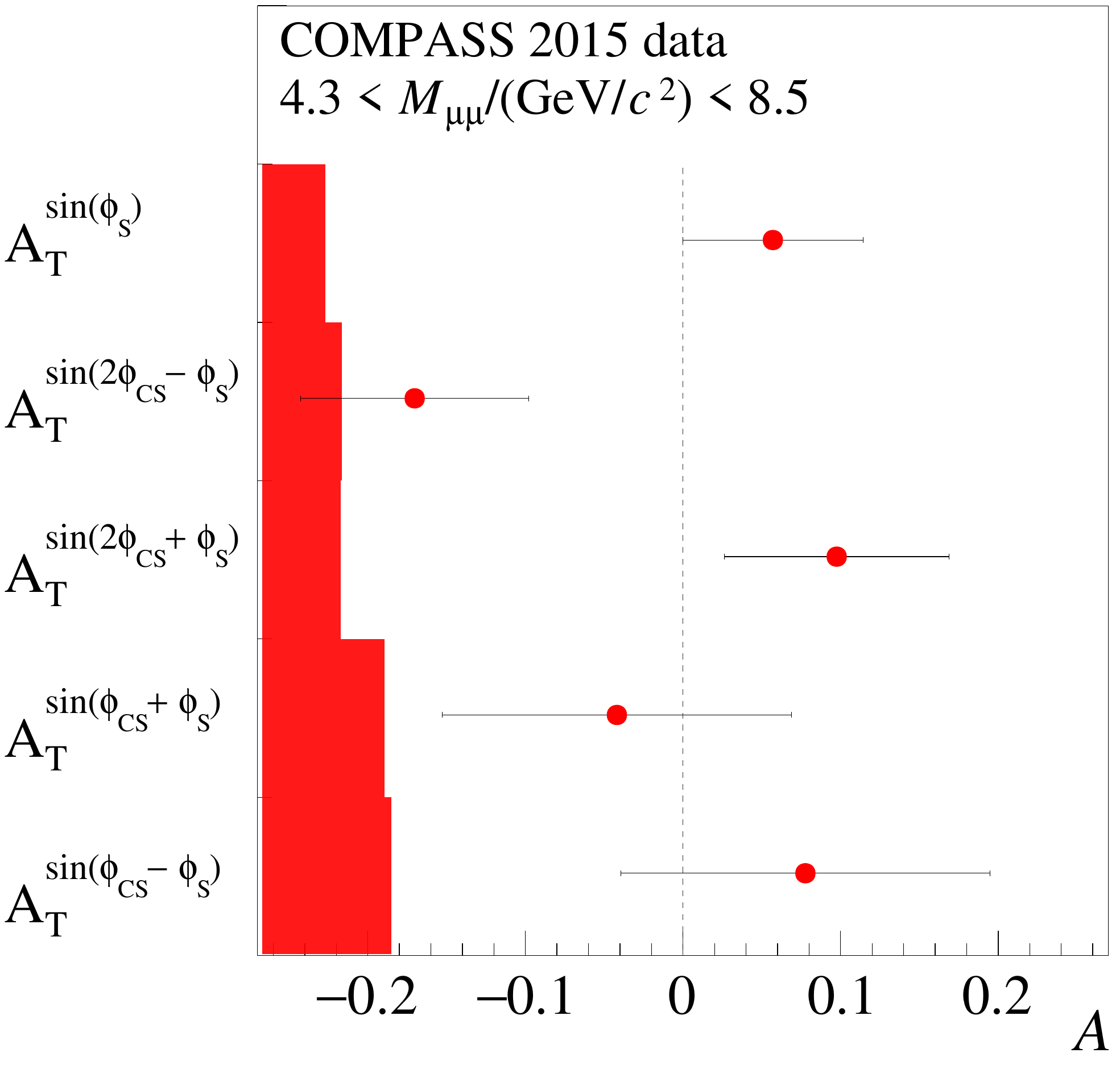}
\end{subfigure}
\caption{Left panel: Extracted Drell-Yan TSAs. Right panel: Mean Drell-Yan TSAs.}
\label{fig:TSAs}
\end{figure}

All five Drell-Yan TSAs (\textit{twist-2} TSAs, $A_T^{\sin\phiS}$, $A_T^{\sin(2\phiCS-\phiS)}$ and $A_T^{\sin(2\phiCS+\phiS)}$ and \textit{subleading-twist} TSAs, $A_T^{\sin(\phiCS-\phiS)}$ and $A_T^{\sin(\phiCS+\phiS)}$) are shown in Fig.~\ref{fig:TSAs} (left panel) as a function of the variables $x_{N}$, $x_{\pi}$, $x_{F}$, $q_{T}$ and \Minv.
Due to relatively large statistical uncertainties, no clear trend is observed for any of the TSAs.
The full set of numerical values for all TSAs from this measurement (including corresponding correlation coefficients)  is available on HepData~\cite{hepdata}.

\begin{wrapfigure}{r}{7.0cm}
\centering
\includegraphics[width=0.45\textwidth]{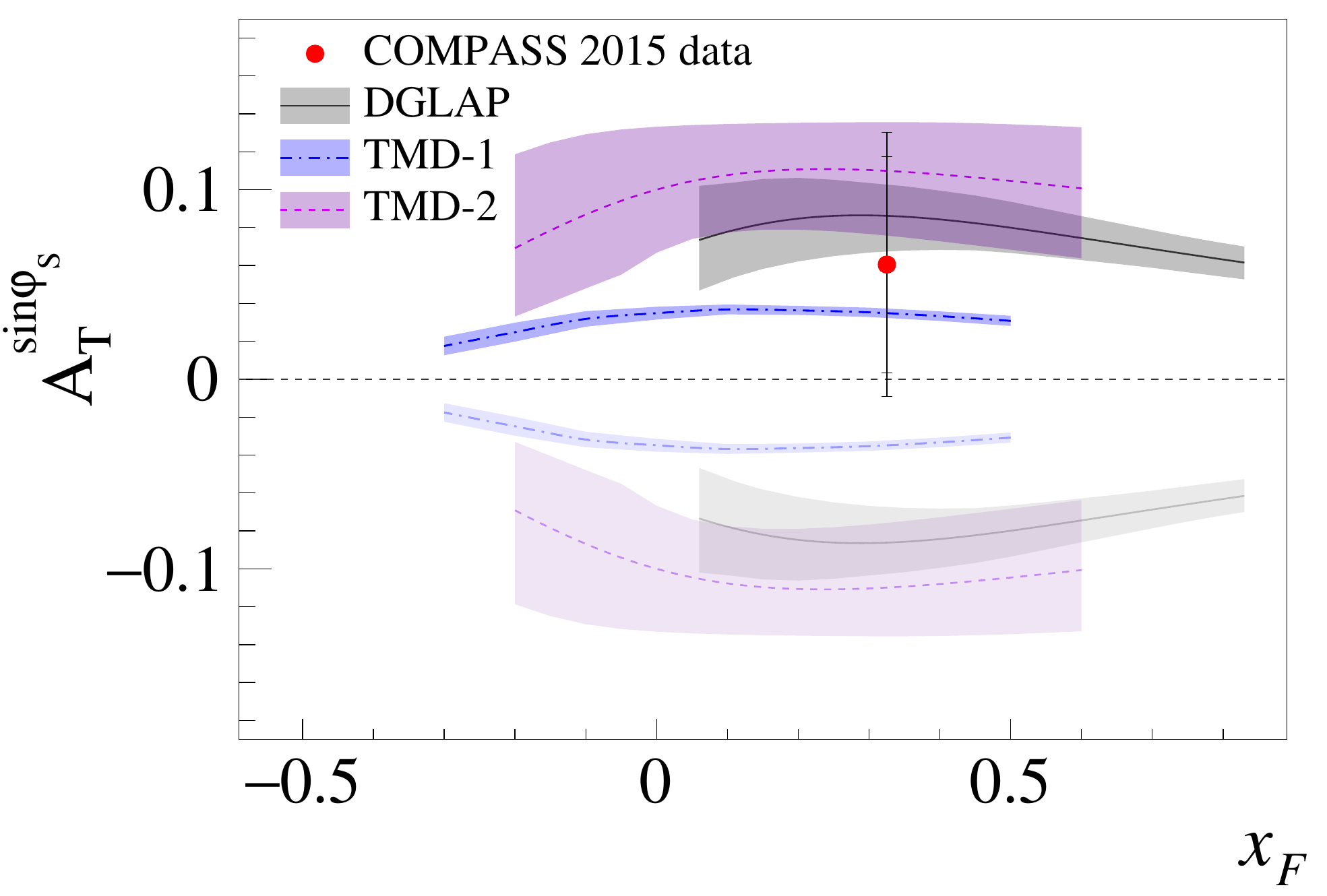}
\caption{The measured mean Sivers asymmetry and the
theoretical  predictions from Refs.~\cite{Anselmino:2016uie} (DGLAP),~\cite{Echevarria:2014xaa} (TMD1)
and~\cite{Sun:2013hua} (TMD2).
The dark-shaded (light-shaded) predictions are evaluated with (without) the sign-change hypothesis}.
\label{fig:Siv_theor}
\end{wrapfigure}

The right panel in Fig.~\ref{fig:TSAs} shows the results for the five extracted TSAs integrated over the entire
kinematic range.
The average Sivers asymmetry $A_T^{\sin\phiS}=0.060\pm0.057(stat.)\pm0.040(sys.)$ is found to be above zero at about one standard deviation of the total uncertainty.
In Fig.~\ref{fig:Siv_theor}, it is compared with recent theoretical predictions from Refs.~\cite{Anselmino:2016uie,Echevarria:2014xaa,Sun:2013hua} that are based on standard DGLAP and two different TMD evolution approaches. 
The positive sign of these theoretical predictions for the DY Sivers asymmetry was obtained by using the sign-change hypothesis for the Sivers TMD PDFs, and the numerical values are based on a fit of SIDIS data for the Sivers TSA~\cite{Airapetian:2009ae,Adolph:2012sp}.
%
%
The figure shows that this first measurement of the DY Sivers asymmetry is consistent with the predicted change of sign for the Sivers function.

The average value for the TSA  $A_T^{\sin(2\phiCS-\phiS)}$ is measured to be below zero with a significance of about two standard deviations. The obtained magnitude of the asymmetry is in agreement with the model calculations of Ref.~\cite{Sissakian:2010zza} and can be used to study the universality of the nucleon transversity function. The TSA $A_T^{\sin(2\phiCS+\phiS)}$, which is related to the nucleon pretzelosity TMD PDFs, is measured to be above zero with a significance of about one standard deviation. Since both $A_T^{\sin(2\phiCS-\phiS)}$ and $A_T^{\sin(2\phiCS+\phiS)}$ are related to the pion Boer-Mulders PDFs, the obtained results may be used to study this function further and to possibly determine its sign.
They may also be used to test the sign change of the nucleon Boer-Mulders TMD PDFs between SIDIS and DY as predicted by QCD~\cite{Collins:2002kn}, when combined with other past and future SIDIS and DY data related to target-spin-independent Boer-Mulders asymmetries~\cite{Falciano:1986wk,Airapetian:2012yg,Adolph:2014pwc}.

The remaining two asymmetries are the $A_T^{\sin(\phiCS-\phiS)}$ and $A_T^{\sin(\phiCS+\phiS)}$) \textit{subleading-twist} TSAs. Both amplitudes are found to be compatible with zero (see Fig.~\ref{fig:TSAs}), which can be attributed to the subleading nature of the effects and corresponding dynamic suppressions.

Presented results are the first and the only currently available data on the transverse-spin-dependent azimuthal asymmetries in the Drell-Yan process. In 2018, COMPASS will continue the measurements of the polarized Drell-Yan process for another data-taking year. This will considerably improve the statistical precision of the Sivers and other azimuthal DY TSAs presented in this Letter. Combined with available SIDIS data, COMPASS DY results serve as a crucial input for study of universality of TMD PDFs and for general understanding of the transverse spin structure of the nucleon.

\bibliography{DIS2017_DY_biblio}{}

\end{document}